\documentclass[journal=apchd5,communication=article]{achemso}

\usepackage[version=3]{mhchem} 
\usepackage[T1]{fontenc}       
\usepackage{caption}
\usepackage{float}
\usepackage{geometry}
\usepackage{natbib}
\usepackage{setspace}
\usepackage{xkeyval}

\author{Ata Chizari}
\affiliation{Department of Electrical Engineering, Sharif University of Technology, Tehran, Iran}
\author{Sajjad AbdollahRamezani}
\affiliation{Department of Electrical Engineering, Sharif University of Technology, Tehran, Iran}
\author{Mohammad Vahid Jamali}
\affiliation{Department of Electrical Engineering, Sharif University of Technology, Tehran, Iran}
\author{\\Jawad A. Salehi}
\affiliation{Department of Electrical Engineering, Sharif University of Technology, Tehran, Iran}
\email{jasalehi@sharif.edu}

\title[An \textsf{achemso} demo]
{Analog Optical Computing Based on Dielectric Meta-reflect-array}

\begin{document}


\begin{abstract}
In this paper, we realize the concept of analog computing using an array of engineered gradient dielectric meta-reflect-array. The proposed configuration consists of individual subwavelength silicon nanobricks in combination with fused silica spacer and silver ground plane realizing a reflection beam with full phase coverage $2\pi$ degrees as well as amplitude range $0$ to $1$. Spectrally overlapping electric and magnetic dipole resonances, such high-index dielectric metasurfaces can locally and independently manipulate the amplitude and phase of the incident electromagnetic wave. This practically feasible structure overcomes substantial limitations imposed by plasmonic metasurfaces such as absorption losses and low polarization conversion efficiency in the visible range. Using such CMOS-compatible and easily integrable platforms promises highly efficient ultrathin planar wave-based computing systems which circumvent the drawbacks of conventional bulky lens-based signal processors. Based on these key properties and general concept of spatial Fourier transformation, we design and realize broadband mathematical operators such as differentiator and integrator in the telecommunication wavelengths.
\end{abstract}

Manipulating the transmission of light has garnered particular attention from ancient times. Conventional optical components such as dielectric lenses rely on the accumulation phase delay during wave propagation to tailor the impinging light wavefront to a desired path \cite{yu2014flat}. However, such bulky geometries suffer from substantial restrictions such as non-negligible reflection loss, spherical aberration, and diffraction limit \cite{abdollahramezani2015analog}. Recently, significant researches have been carried out in the area of optical metamaterials, defined broadly as artificially engineered configurations designed in the subwavelength regime \cite{zheludev2012metamaterials}. Different from the traditional optical elements, such structures open up the exciting possibility of tailoring wave behaviors by providing access to a broader range for constituent material parameters such as permittivity, permeability, refractive index, chirality, and nonlinear susceptibility. Based on the resonant interaction between incoming light and inclusions, numerous exotic phenomena and functionalities have been exhibited such as optical cloaking devices \cite{alu2005achieving}, metamaterial nanocircuits \cite{engheta2007circuits}, and more. However, the general usage of metamaterials precluded by three-dimensional ($3$D) fabrication complexity and inherent high losses \cite{yu2014flat, kildishev2013planar}.

Metasurfaces, or metamaterials of reduced dimensionality, have been of great interest due to their benefits of possessing simple $2$D fabrication procedure, smaller physical footprint, and lower losses compared to their $3$D bulky counterparts \cite{kildishev2013planar, zheludev2012metamaterials}. Plasmonic metasurfaces consist of a dense arrangement of subwavelength resonant antennas providing flexibility to manipulate the amplitude, phase, and polarization responses of the scattered light. By configuring the geometry dimensions, shape, period, orientation, and material of the constituent structure, metasurfaces function as interface discontinuities that locally, independently, and abruptly molding incoming light \cite{qin2016hybrid, farmahini2013birefringent}. As a result, planar analogs of conventional optical components have been emerged such as ultrathin focusing lenses \cite{aieta2015multiwavelength}, holograms \cite{zheng2015metasurface}, and beam forming \cite{farmahini2013birefringent}.

It is worth mentioning that a single layer of transversally inhomogeneous metasurface only allows for full $2\pi$-phase control of the cross-polarized light component due to their Lorentzian-shaped polarizability, and hence drastically limits the efficiency of structure to  $25$\% \cite{zhang2016advances}. Although by using two cascaded surfaces the transmission level might be improved, the accessible range of phase values would intrinsically be limited \cite{monticone2013full}. By employing a transmit-array of several plasmonic metasurfaces, a moderate efficiency of $25-50$\% could be achieved; however, such structures are quite complex to be fabricated at near-infrared and visible frequency range \cite{pfeiffer2013cascaded,abdollahramezani2015beam}. 

In order to boost up the manipulation efficiency up to $80$\%, an easy-to-fabricate approach that works in reflection has been proposed utilizing a periodic array of nanoscale metallic nanobrick on top of a subwavelength thin dielectric spacer settled on an optically thick metal ground \cite{pors2013broadband}. Based on mode coupling, this metal-insulator-metal (MIM) structure supports a magnetic resonance with strong magnetic field coupled in the dielectric spacer due to intense coupling between dipole-like localized surface plasmon resonance atop and its image. Although this design avoids complexity, it is still compromised by strong nonradiative Ohmic and polarization-conversion losses at optical frequencies \cite{lin2014dielectric}. 

Metasurfaces composed of dielectric resonators are promising candidates to overcome the fundamental challenge of high energy dissipation due to metallic losses. Interacting with incident light, high-index nanoresonators support both magnetic and electric resonances \cite{staude2013tailoring}. Therefore, they behave like magnetic dipole (first Mie resonance) possessing circular displacement current which allows an enhanced magnetic field at the center of each resonator. Moreover, owing to the accumulated charges at the end of resonator, an electric dipole (second Mie resonance) is appeared leading to a strongly dominated electric field in the heart of nanoparticle \cite{jahani2016all}. Following a characteristic resonant dispersion in the linear response regime, both negative and positive effective constitutive parameters, namely permittivity and permeability, could be achieved around the electric and magnetic resonances, respectively. Recent researches have also shown the ability of high-index nanoresonators to manipulate phase and polarization of light waves leading to applications such as flat lenses \cite{yu2015high}, polarizers \cite{arbabi2015dielectric}, and vortex beam generation \cite{yang2014dielectric}.

Recently, optical analog computing based on optical metamaterial and plasmonic metasurfaces has emerged as a promising candidate for real-time and parallel continuous data processing \cite{silva2014performing, pors2014analog, farmahini2013metasurfaces,youssefi2016brewster}. To overcome all the aforementioned challenges related to such structures, we propose an easy-to-fabricate dielectric meta-reflect-array (MRA) which performs mathematical operations by using engineered nanobricks on top of a silica spacer and a silver substrate. Supporting both electric and magnetic dipolar Mie-type modes as well as multiple reflections within the spacer layer enable us to fully manipulate not only amplitude but also phase profile of the reflected cross-polarized light spatially by varying nanoresonator dimensions.

\begin{figure} 
\centering
\includegraphics[trim=0cm 8.6cm 5cm 0cm,width=16.6cm,clip]{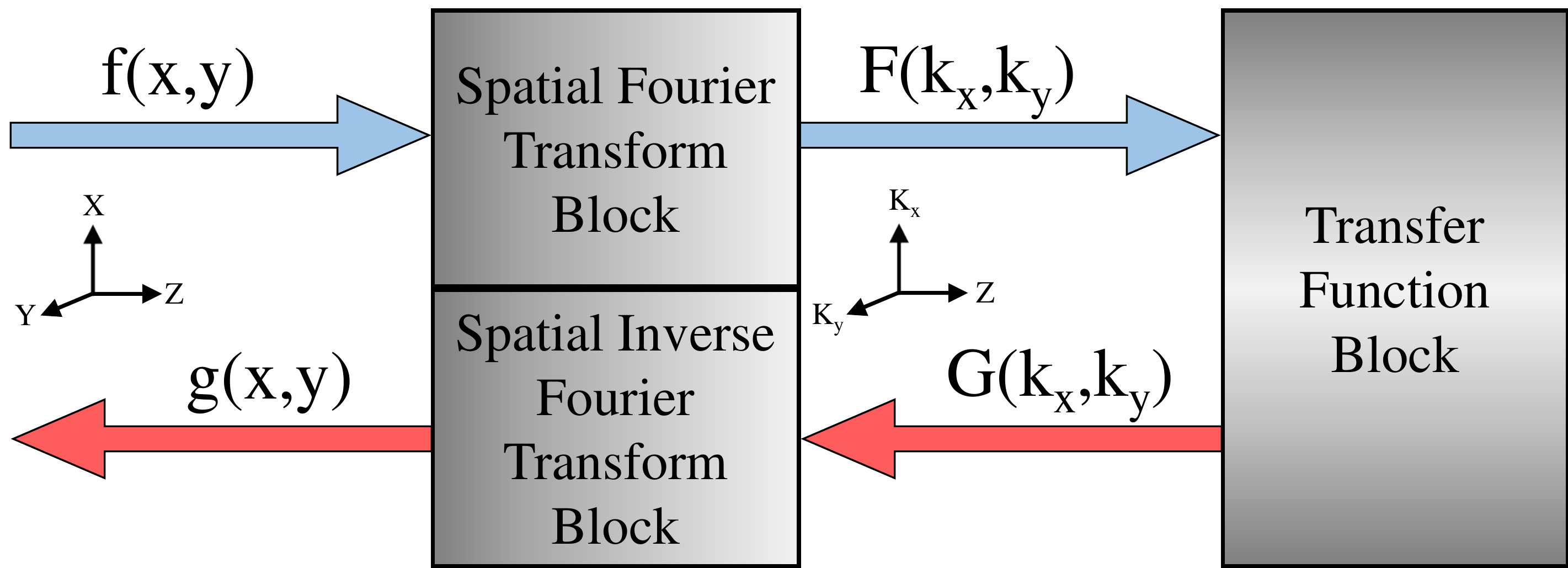}\\
\captionsetup{justification=justified}
\caption{Sketch of mathematical operations based on general concept of spatial Fourier transformation for the proposed system.}\label{flowdiagram}
\end{figure}

Accordingly, our approach for realizing mathematical operations based on spatial Fourier transformation is schematically illustrated in Fig.~\ref{flowdiagram}. Consider a linear space-invariant system in which $f(x,y)$ and $g(x,y)$ are the arbitrary input and corresponding output functions, respectively . These two functions relate to each other via the system's $2$D impulse response, $h(x,y)$, according to the following linear convolution \cite{silva2014performing}:
\begin{align} 
g(x,y)&=h(x,y){\ast}f(x,y)\nonumber\\
		 &={\int}{\int}h(x-x',y-y')f(x',y')\textnormal{d}x'\textnormal{d}y'
\label{equ1}
\end{align}
Eq.~1 can be represented in the spatial Fourier domain as:
\begin{equation} 
G(k_{x},k_{y})=H(k_{x},k_{y})F(k_{x},k_{y})
\label{equ2}
\end{equation}
where $G(k_{x},k_{y})$, $H(k_{x},k_{y})$, and $F(k_{x},k_{y})$ are the Fourier transform of their counterparts in Eq.~1, respectively, and $(k_{x},k_{y})$ denotes the $2$D variables in the Fourier space. 

\begin{figure}
	\centering
	\includegraphics[trim=0cm 0.1cm 0cm 0cm,width=16.6cm,clip]{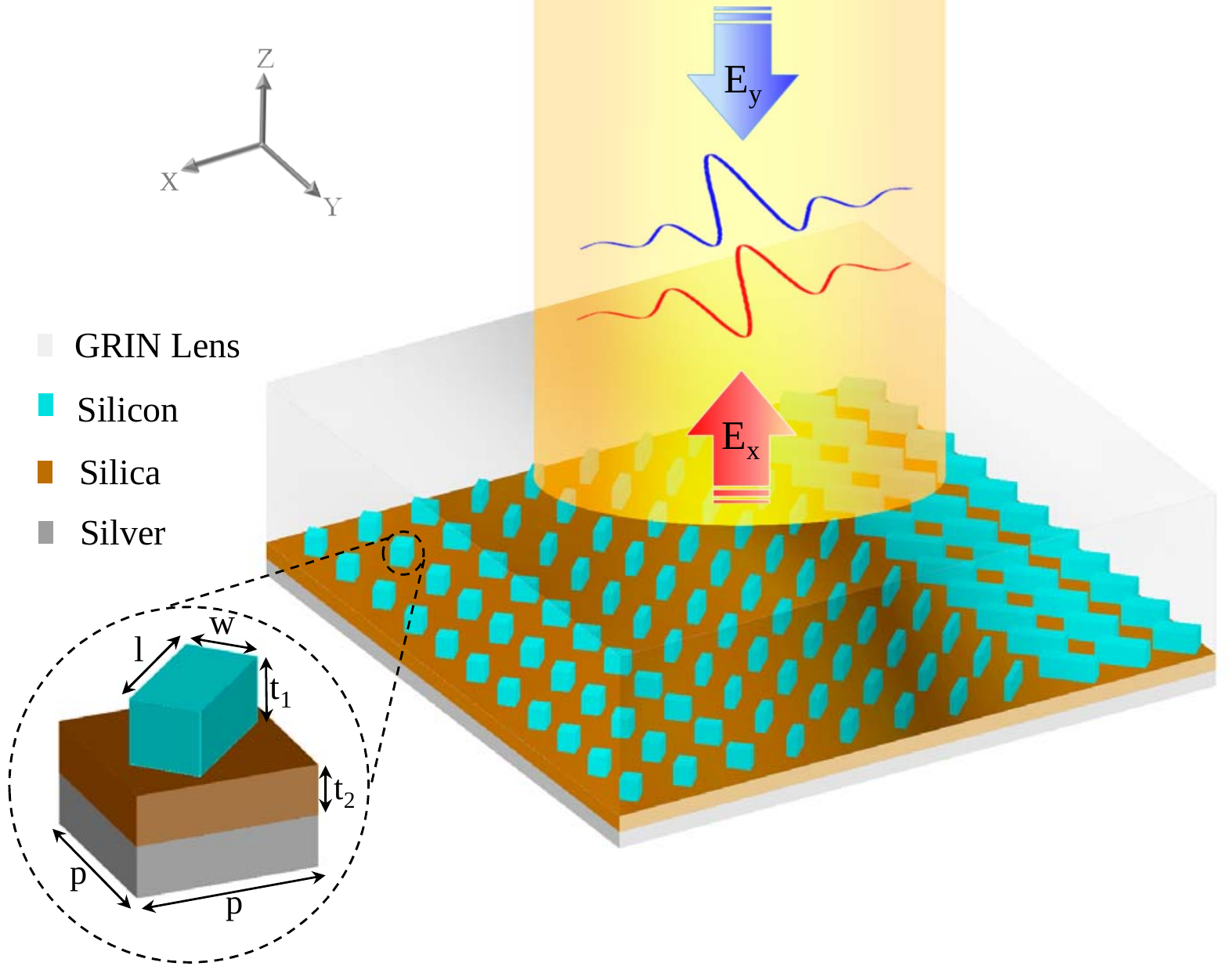}\\
	\caption{Schematic demonstration of the proposed GRIN/MRA structure and its constitutive unit cell.}\label{Schematic}
\end{figure}

For our proposed system shown in Fig.~\ref{Schematic}, $f(x,y)$ and $g(x,y)$ are electric field profiles of the incident and cross-polarized reflected monochromatic wave signals, i.e., $E_{y}(x,y)$ and $E_{x}(x,y)$, respectively. $H(k_{x},k_{y})$ is the transfer function associated with the desired mathematical operation which can be considered as position-dependent reflection coefficient $R(k_{x},k_{y})$ due to the reflective nature of structure. Although realizing $2$D complex mathematical operations by engineering unit cells along $x$ and $y$ directions is not challenging, we, here, limit our discussion to a $1$D system which is symmetric along $y$ axis.  Consequently, Eq.~2 could be expressed as:
\begin{equation} 
E_{x}(x)=\textnormal{IFT}\lbrace{R(k_{x})\textnormal{FT}[E_{y}(x)]}\rbrace
\label{equ3}
\end{equation}
where (I)FT means (inverse) Fourier transform. It is worth mentioning that the Fourier variable $k_{x}$ is in the same domain of $x$; hence, the real-space coordinate $x$ plays the role of $k_{x}$ \cite{silva2014performing}.  

While Fourier transform can readily be realized by employing a graded index (GRIN) lens \cite{silva2014performing}, performing inverse Fourier transform is not possible by natural materials. However, based on the well-known formula $\textnormal{FT}\lbrace{\textnormal{FT}[E_{x}(x)]}\rbrace\propto{E_{x}(-x)}$ implying that the output will be proportional to the mirror image of the desired output field distribution $E_{x}(x)$, we employ the same GRIN lens, which has been used as the FT block in the input, instead of the IFT block in the output. Furthermore, in order to implement the appropriate transfer function, or equivalently the reflection coefficient, a well-designed structured high-index dielectric metasurface is being applied. Actually, each unit cell functions as a nanoscale spatial light modulator which locally and independently covers the full $2\pi$ phase range while at the same time reaches strong modulation in the reflection amplitude, i.e., span [$0,1$] .

\begin{figure} 
\centering
\includegraphics[trim=11.2cm 10cm 0cm 0cm,width=16.6cm,clip]{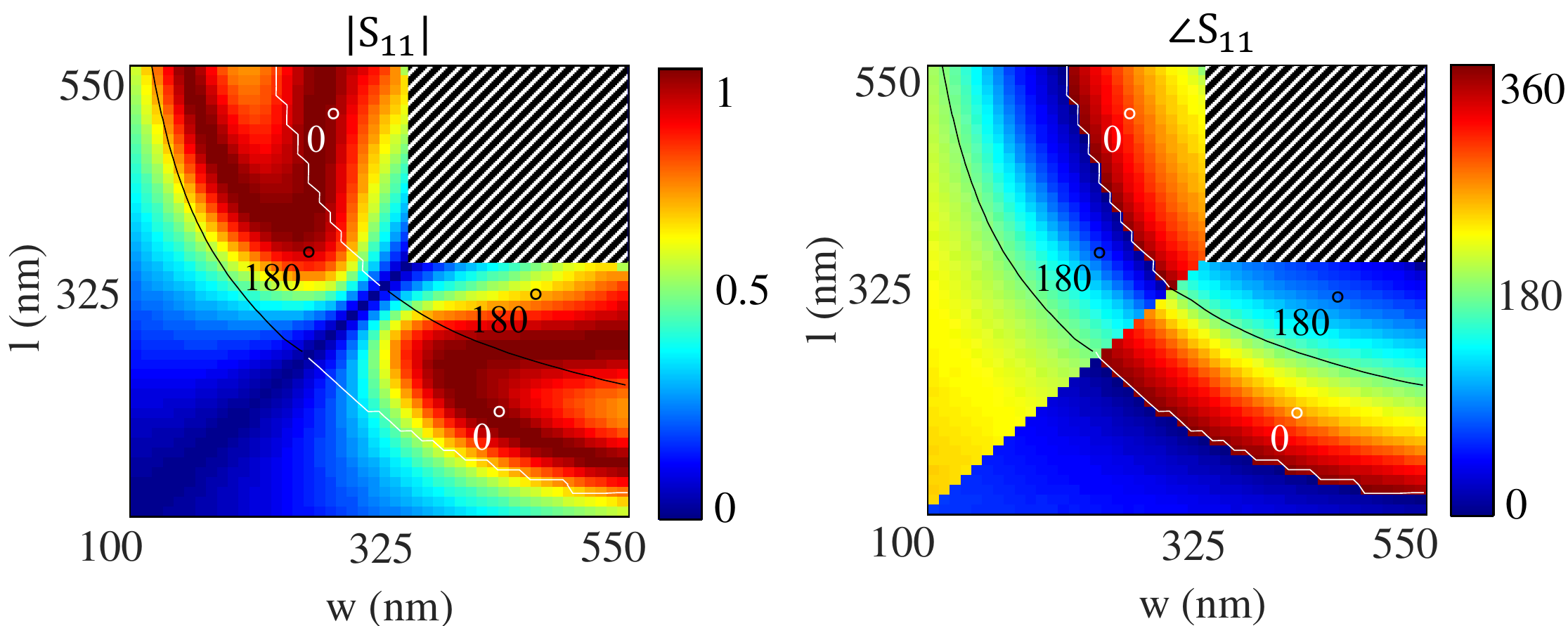}\\
\captionsetup{justification=justified}
\caption{ Simulated (a) amplitude and (b) corresponding phase, profiles of reflection coefficients for the proposed structure versus width and length of a nanobrick for $p=650$ nm, $t_{1}=380$ nm, and $t_{2}=200$ nm.}\label{MATLAB}
\end{figure}

To demonstrate such independent tunning, the amplitude and phase of reflection coefficient versus the width ($w$) and length ($l$) of nanobrick are depicted in Fig.~\ref{MATLAB}, where a normally beam, linearly polarized along $y$-direction, at $1550$ nm is impinging onto the unit cell. Full-wave electromagnetic simulations are carried out using CST Microwave Studio based on finite-element frequency-domain (FEFD) technique. Notably, the reflection amplitude and phase are obtained from the $S_{11}$ parameter. Each unit cell consists of a silicon nanobrick (with thickness $t_{1}=380$ nm and refractive index $n_{silicon}=3.7$) on top of a fused silica spacer (with thickness $t_{2}=200$ nm and refractive index $n_{silica}=1.48$) and an optically thick silver substrate. The periodicity in $x$ and $y$ directions is $p=650$ nm. Using an optically thick silver layer simplifies numerical simulations; however, an almost same optical response can be achieved with a subwavelength thickness layer around $100$ nm. Hence, the overall thickness of the MRA is still being subwavelength.

The Fabry-Per\'ot-like cavity appears due to the presence of silver substrate and nanobrick array causes the consequent interference of polarization couplings in the multireflection process to enhance and reduce the cross-polarized and co-polarized reflected fields, respectively. It is notable that multiple reflections within the spacer layer also lead to smoothed phase variation across the resonance \cite{grady2013terahertz}. These effects as well as Mie resonances discussed before result in any desirable transmission phase and amplitude profile achieved by properly structuring each nanobrick (see Fig.~\ref{MATLAB}).

The first calculus-MRA to be designed is differentiator. According to the Fourier transform principles, the $n^{th}$ spatial differentiation of a $1$D function is related to its first Fourier transform as:
\begin{equation} 
\dfrac{\textnormal{d}^{n}(f(x))}{\textnormal{d}x^{n}}=\textnormal{IFT}\lbrace(ik_{x})^{n}\textnormal{FT}(f(x))\rbrace
\label{equ4}
\end{equation}
Here, the term $(ik_{x})^{n}$ or equivalently $(ix)^{n}$, as explained before, plays the role of transfer function or reflection coefficient, which should be implemented by a well-structured MRA. Since MRA is a passive media with finite lateral dimension limited by $D$, the reflection coefficient amplitude should be remained below unity across the transverse plane. To do so, the desired transfer function becomes $R(x)\propto(ix/(D/2))^{n}$.

\begin{figure} 
\centering
\includegraphics[trim=0cm 0.1cm 0cm 0cm,width=16.6cm,clip]{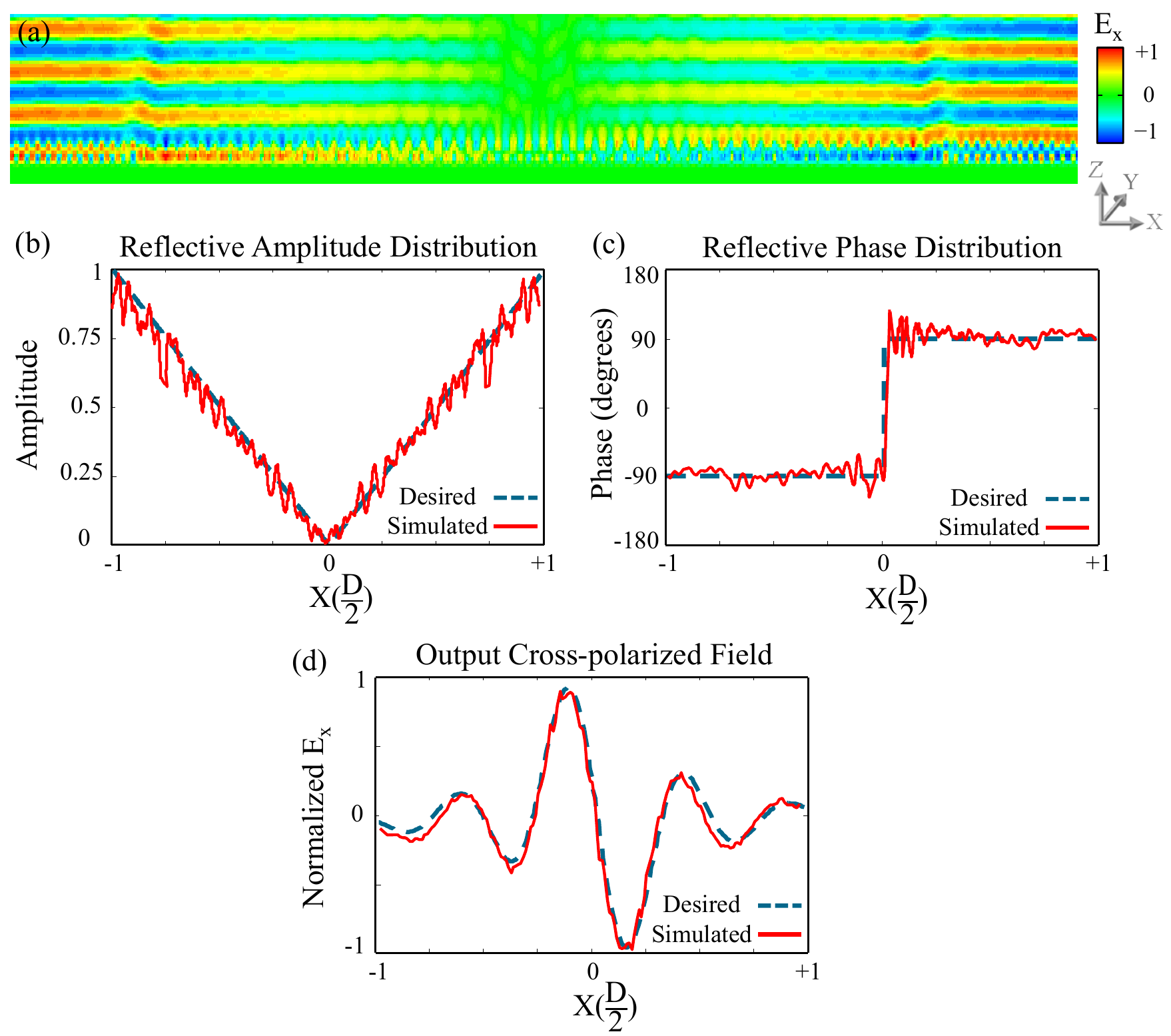}\\
\captionsetup{justification=justified}
\caption{(a) Snapshot of reflected electric field distribution ($E_{x}$). Comparison of the theoretical and simulation results of (b) amplitude, (c) phase, and (d) output cross-polarized electric field of the reflected beam for the first-order derivation functionality.}\label{Diff1}
\end{figure}

Fig.~\ref{Diff1}(b) and Fig.~\ref{Diff1}(c) illustrate the desired amplitude and phase distributions of the first-order derivative transfer function. According to Fig.~\ref{MATLAB}(a) and Fig.~\ref{MATLAB}(b), by properly tailoring the width and length of each nanobrick along the lateral direction, the transverse amplitude and phase distribution of the transfer function could readily be implemented. To this end, the corresponding dimensions of each nanoresonator have to be chosen along the black and white lines where the reflection amplitude can be tuned from $0$ to $1$ while the phase difference is kept as $180^{\circ}$. In order to validate the performance of the proposed MRA, a linear $y$-polarized plane wave impinges toward the structure. As shown in Fig.~\ref{Diff1}(b) and Fig.~\ref{Diff1}(c), calculated electric field distribution of reflected cross-polarized beam is in good agreement with the desired transfer function.

\begin{figure} 
\centering
\includegraphics[trim=0cm 0.15cm 0cm 0cm,width=16.6cm,clip]{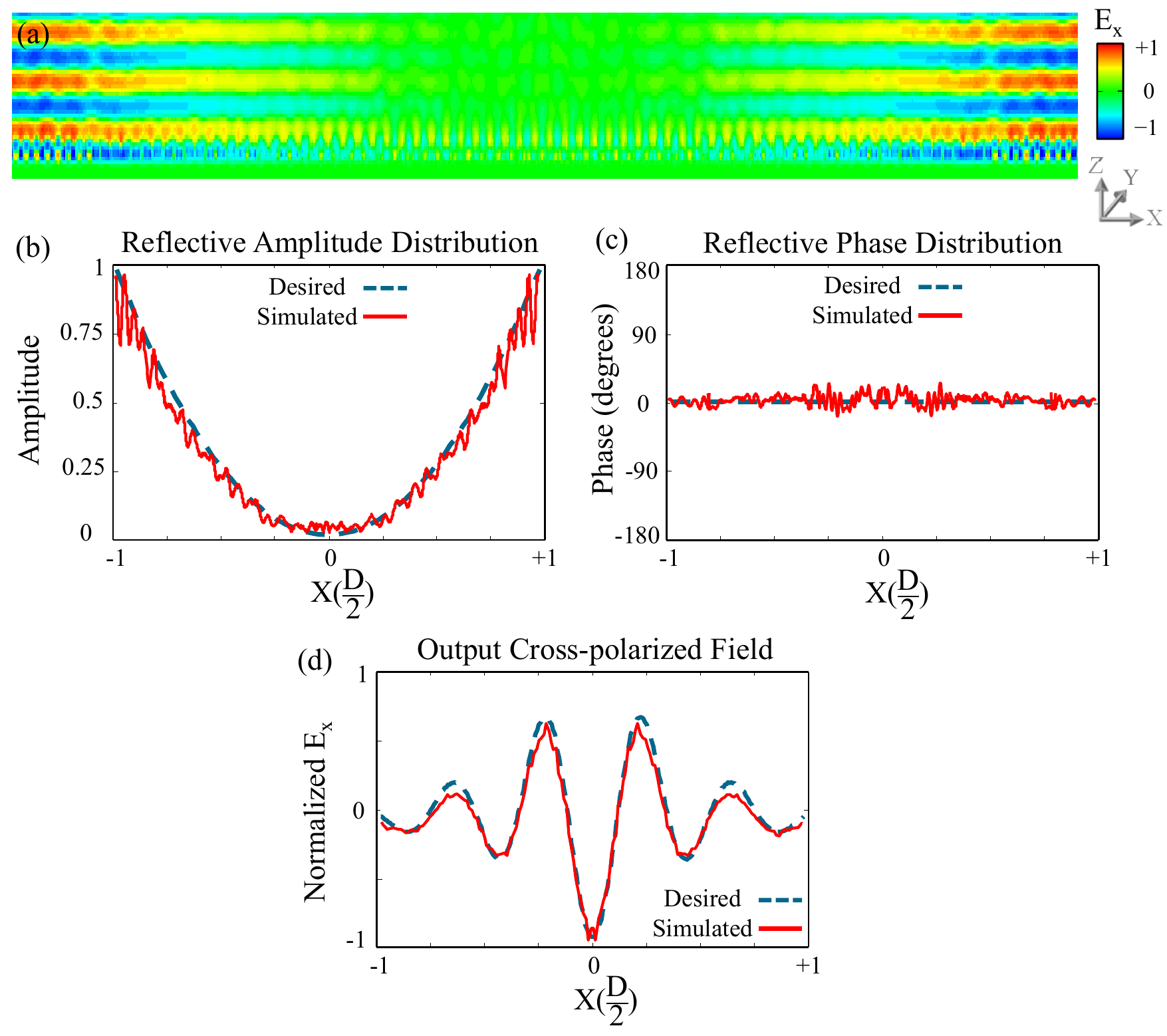}\\
\captionsetup{justification=justified}
\caption{(a) Snapshot of reflected electric field distribution ($E_{x}$). Comparison of the theoretical and simulation results of (b) amplitude, (c) phase, and (d) output cross-polarized electric field of the reflected beam for the second-order derivation functionality.}\label{Diff2}
\end{figure}

As another design, we implement second-order differentiator with desired transfer function $R(x)\propto-(x/(D/2))^{2}$. The amplitude and phase profiles of $R(x)$ are plotted in Fig.~\ref{Diff2}(b) and Fig.~\ref{Diff2}(c), respectively. It is obvious that corresponding amplitude and phase of reflection coefficient have a quadratic distribution and constant phase, respectively. To this end, $w$ and $l$ for each nanobrick should be chosen along the white solid lines in Fig.~\ref{MATLAB}(b) to cover all desired amplitude values between $0$ and $1$ with constant phase of $0^{\circ}$. Based on Fig.~\ref{Diff2}(b) and Fig.~\ref{Diff2}(c), the output function of our numerical simulation quite reasonably resembles the expected output. 

\begin{figure}
	\centering
	\includegraphics[trim=.1cm .1cm .2cm 0cm,width=16.6cm,clip]{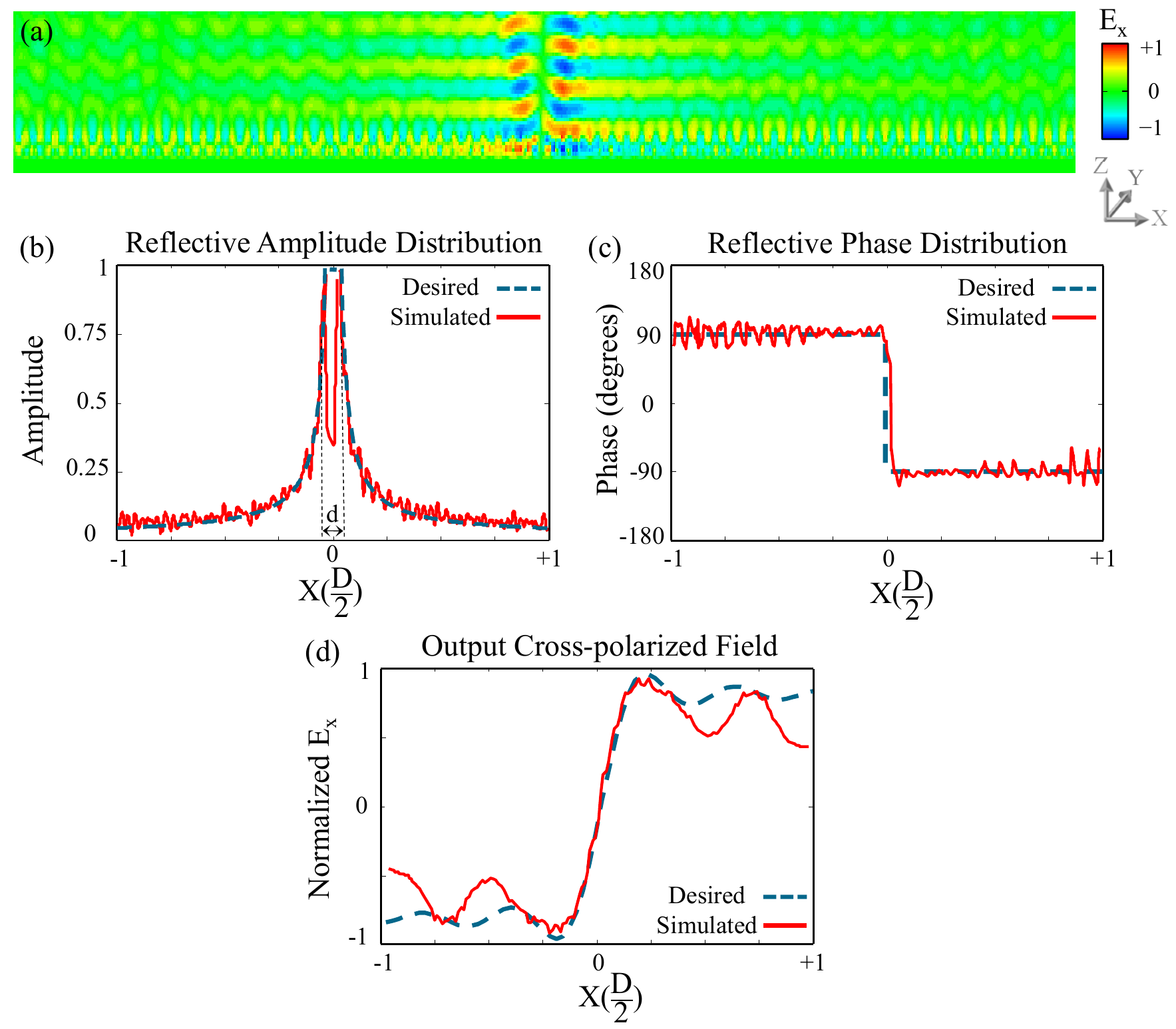}\\
	\captionsetup{justification=justified}
	\caption{(a) Snapshot of reflected electric field distribution ($E_{x}$). Comparison of the theoretical and simulation results of (b) amplitude, (c) phase, and (d) output cross-polarized electric field of the reflected beam for the first-order integration functionality.}\label{Int}
\end{figure}

In the case of integration functionality, the transfer function becomes $R(x)\propto(ix)^{-1}$. Due to the infinite values appear for $R(x)$ in the vicinity of origin, i.e., $x=0$, we truncate the required transfer function at distance $d$ and assume the amplitude profile to be unity for all points within this range to keep away gain requirements. Therefore, approximated transfer function follows:
\begin{equation}
R(x)=\begin{cases}
\hspace{3mm}1\hspace{6mm}, & \text{if $\vert{x}\vert<d$}\\
(ix)^{-1}, & \text{if $\vert{x}\vert>d$}
\end{cases}
\label{equ12}
\end{equation}
where $d$ is an arbitrary parameter chosen as $d=D/20$ for this case \cite{silva2014performing}. Numerical simulation results depicted in Fig.~\ref{Int}(b) and Fig.~\ref{Int}(c) verify reasonable accordance between analytical solution of the desired transfer function of the first-order integration and the reflected electric field distribution.     

To investigate the performance of spatial derivation and integration functionalities, a normally $y$-polarized light with Sinc-shaped electric field $f(x)=\textnormal{Sinc}(x/6.8)$ is incident onto the GRIN/MRA configuration and the cross-polarized reflected electric field distribution for the first-order derivative, second-order derivative, and integrator are demonstrated in Fig.~\ref{Diff1}(d), Fig.~\ref{Diff2}(d), and Fig.~\ref{Int}(d), respectively. Comparing the simulation results with analytical ones confirms the acceptable agreement.

In summary, we have proposed and designed mathematical operators using a GRIN/MRA configuration at telecommunication wavelengths. The proposed meta-reflect-array consists of high-index dielectric nanoresonators which can fully manipulate both amplitude and phase of reflected cross-polarized field locally and independently. Suppoting overlapped electric and magnetic resonances as well as high polarization conversion efficiency, dielectric metasurfaces are superior to their lossy plasmonic counterparts. However, slightly large unit cell size and consequently higher coupling of adjacent cells might be challenging in higher frequencies. Good agreement between simulated and theoretical results confirms the effectiveness of the proposed configuration in realizing mathematical operators, namely derivators and integrator. Such appealing finding may lead to large improvements in image processing, complex wave shaping, and analog computing at the hardware level.

\providecommand*\mcitethebibliography{\thebibliography}
\csname @ifundefined\endcsname{endmcitethebibliography}
  {\let\endmcitethebibliography\endthebibliography}{}


\newpage

\end{document}